\documentclass[conference]{IEEEtran}
\IEEEoverridecommandlockouts
\usepackage{cite}
\usepackage{amsmath,amssymb,amsfonts}
\usepackage{algorithmic}
\usepackage{graphicx}
\usepackage{textcomp}
\usepackage{xcolor}
\usepackage{booktabs}
\usepackage{tabularx}
\usepackage{makecell}
\usepackage{array}   
\usepackage{threeparttable}
\usepackage{url}           

\def\BibTeX{{\rm B\kern-.05em{\sc i\kern-.025em b}\kern-.08em
    T\kern-.1667em\lower.7ex\hbox{E}\kern-.125emX}}
\begin{document}

\title{Understanding Human Daily Experience Through Continuous Sensing: ETRI Lifelog Dataset 2024
\thanks{This work was supported by Electronics and Telecommunications Research Institute (ETRI) grant funded by
the Korean government. [25ZB1100, Core Technology Research for Self-Improving Integrated Artificial Intelligence System].}
}

\author{
\begin{minipage}[t]{\linewidth}\centering
Se Won Oh, Hyuntae Jeong, Seungeun Chung, Jeong Mook Lim,\\
Kyoung Ju Noh, Sunkyung Lee, Gyuwon Jung \\
\textit{Electronics and Telecommunications Research Institute} \\
Daejeon, South Korea \\
\{sewonoh, htjeong, schung, jmlim21, kjnoh, sklee2014, gwjung\}@etri.re.kr
\end{minipage}
}

\maketitle

\begin{abstract}
Improving human health and well-being requires an accurate and effective understanding of an individual's physical and mental state throughout daily life. To support this goal, we utilized smartphones, smartwatches, and sleep sensors to collect data passively and continuously for 24 hours a day, with minimal interference to participants' usual behavior, enabling us to gather quantitative data on daily behaviors and sleep activities across multiple days. Additionally, we gathered subjective self-reports of participants' fatigue, stress, and sleep quality through surveys conducted immediately before and after sleep. This comprehensive lifelog dataset is expected to provide a foundational resource for exploring meaningful insights into human daily life and lifestyle patterns, and a portion of the data has been anonymized and made publicly available for further research. In this paper, we introduce the ETRI Lifelog Dataset 2024, detailing its structure and presenting potential applications, such as using machine learning models to predict sleep quality and stress.
\end{abstract}


\section{Introduction}
Human daily life consists of a complex interrelation of different activities and physiological states, spanning daytime behavior and nighttime sleep. To better understand the intricate nature of human behavior in daily life, it is important to systematically collect and analyze comprehensive lifelogs from a variety of sensor devices and user records~\cite{chan2024capture, nepal2024capturing}. Such continuous and well-quantified data are essential for reliably assessing an individual’s physical and mental states throughout the day, thereby supporting efforts to improve human health and well-being. Advances in sensor technology and data collection methods have enabled tracking of human activity and physiological signals in controlled environments or during specific tasks. In addition, mobile and wearable devices such as smartphones and smartwatches are widely used for health monitoring and behavior tracking~\cite{huhn2022impact, niknejad2020comprehensive, peake2018critical, hardjianto2025graph}.

However, despite these advances, capturing continuous and naturalistic daily experiences in real-world settings without disrupting individuals' routines remains challenging~\cite{hicks2019best, vaizman2018context}. Most existing studies have focused on short-term measurements or controlled laboratory environments, providing only a limited perspective on the complexity of human life. For example, although several studies have used smartphones and smartwatches to track human activity, few have comprehensively monitored a 24-hour cycle of daily life, including sleep, in natural environments over a long period of time~\cite{yfantidou2022lifesnaps, chung2022real, wang2018tracking}.

This gap underscores the need for a dataset that seamlessly integrates continuous, passive measurements of daytime behavior and nighttime sleep patterns to reveal the underlying connections between daily activities, sleep quality, and subjective indicators of well-being such as fatigue, mood, and stress. To address this need, our study collected continuous, unobtrusive data using smartphones, smartwatches, and under-mattress sleep sensors~\cite{lim2023digital}. Subjective self-reports were also taken immediately before and after sleep to capture a holistic view of daily life.

We extended previous efforts, such as the 2020 and 2023 datasets~\cite{chung2022real, etrilifelog2020, oh2024sensor, etrilifelog2024}, but further addressed some key limitations. In this new dataset, we expanded the types of sensor data collected (from 9 to 12 types) while minimizing device load and battery consumption by adjusting the data collection intervals and excluding raw acceleration data. This approach ensured a comprehensive dataset suitable for long-term observation of natural human behavior.

As with the previous ICTC, where our lifelog dataset led to diverse research outcomes~\cite{ictc2024_1, ictc2024_2, ictc2024_3, ictc2024_4, ictc2024_5, ictc2024_6, ictc2024_7, ictc2024_8, ictc2024_9, ictc2024_10, ictc2024_11, ictc2024_12}, the ETRI Lifelog Dataset 2024~\cite{etrilifelog2024} introduced in this paper would provide a foundational resource for exploring meaningful insights into human routines and life patterns. For example, it can be used in studies to investigate how daily behaviors and sleep activities relate to subjective and physiological states. We demonstrate a real-world application of this dataset by using machine learning models to predict sleep quality, illustrating its relevance for human-centered health research.

The remainder of this paper is organized as follows: we describe the data collection methods and dataset structure, then present an application involving machine learning models for sleep quality prediction, followed by discussion and the conclusion. This study provides a robust and natural dataset that can support research on human well-being and daily life and promote the development of data-driven health solutions.

\section{Data Collection and Dataset Overview}

This dataset represents a portion of the outcomes from a research project jointly conducted by Electronics and Telecommunications Research Institute (ETRI) and Chungnam National University Hospital (CNUH). It has been curated and released to promote further research and includes a total of 700 days of lifelog data collected from 10 participants who took part in experiments conducted between May and December 2024. Each subject is assigned a unique identifier (e.g., `subject\_id'), and their demographic details are summarized in Table~\ref{tab:demographics_ch2025}. Subject ages are grouped in 10-year increments, while height and weight data are rounded for simplicity. 

\begin{table}[t!]
    \caption{Demographic characteristics of subjects\strut}
    \centering
    \begin{tabular}{c|c|c|c|c}
    \toprule
    subject\_id & Gender & Age Group & Employment & BMI\\
    \midrule
    id01 & female   & 40s & Yes & 23.7\\
    id02 & male     & 50s & Yes & 22.7\\
    id03 & female   & 30s & No & 29.1\\
    id04 & female   & 40s & No & 19.6\\
    id05 & female   & 50s & Yes & 26.3\\
    id06 & male     & 20s & Yes & 27.1\\
    id07 & female   & 40s & Yes & 26.3\\
    id08 & male     & 30s & Yes & 29.3\\
    id09 & male     & 30s & Yes & 25.4\\
    id10 & female   & 30s & Yes & 26.0\\
    \bottomrule
    \end{tabular}
    \label{tab:demographics_ch2025}
\end{table}

During daytime activities, the participants wore smartphones and smartwatches, while nighttime sleep activities were recorded using sleep sensors installed in their personal living spaces. Additionally, immediately before and after sleep, participants recorded survey inputs on fatigue, stress, and sleep quality using an ecological momentary assessment (EMA) approach. 

Table~\ref{tab:data_items} summarizes twelve data items collected using smartphones and smartwatches. The data count of each item represents the total number of data instances. Each item was collected using the participants’ Android smartphones or smartwatches, with data sampling intervals ranging from one to ten minutes. It should be noted that some data items may contain a considerable amount of missing data and noise. Since this dataset was collected in everyday life rather than in a controlled laboratory environment, potential noise may be present. Additionally, sensor measurements and recordings may be interrupted due to device charging or the rebooting process.

\begin{table*}[h!]
    \caption{Summary of data items collected using smartphones and smartwatches}
    \label{tab:data_items}
    \centering
    \begin{minipage}{\textwidth}
    \begin{tabularx}{\textwidth}{*{3}{>{\centering\arraybackslash}X} l}
    \toprule    
    \textbf{Item} & \textbf{Data count} & \textbf{Sampling interval} & \multicolumn{1}{c}{\textbf{Note}} \\
    
    \midrule
    mACStatus & 939,896& 10-min& Indicates whether the smartphone is currently being charged\\ \hline
    mActivity & 961,062& 1-min& Value calculated by the Google Activity Recognition API\footnote{\url{https://developers.google.com/location-context/activity-recognition}}\\ \hline
    mAmbience & 476,577& 2-min& Ambient sound identification labels\footnote{\url{https://research.google.com/audioset/ontology}} and their respective probabilities\\ \hline
    mBle & 21,830& 10-min& Bluetooth devices around individual subject\\ \hline
    mGps & 800,611& 1-min& Multiple GPS coordinates measured within a single minute using the smartphone\\ \hline
    mLight & 96,258& 10-min& Ambient light measured by the smartphone\\ \hline
    mScreenStatus & 939,653& 1-min & Indicates whether the smartphone screen is in use\\ \hline
    mUsageStats & 45,197& 10-min& Indicates which apps were used on the smartphone and for how long\\ \hline
    mWifi & 76,336& 10-min& Wifi devices around individual subject\\ \hline
    
    wHr & 382,918& 1-min& Heart rate readings recorded by the smartwatch\\ \hline
    wLight & 633,741& 1-min& Ambient light measured by the smartwatch\\ \hline
    wPedo & 748,100& 1-min& Step data recorded by the smartwatch\\
    \bottomrule
    \end{tabularx}
    \end{minipage}
\end{table*}

The detailed structure of individual sensor data items is summarized in Table~\ref{tab:detailed_data_items_ch2025}. Each data item is stored as an individual data file and is provided along with the subject identifier and timestamp. All timestamps are recorded based on \textbf{Korea Standard Time (KST)} and are displayed in the \textbf{YYYY-MM-DD HH:MM:SS} format (e.g., 2024-08-01 12:34:56). To protect privacy, certain data items have undergone minimal anonymization. For example, GPS latitude and longitude data are provided as relative coordinates.

\begin{table*}[h!]
    \caption{Summary of individual data item names (column names), data types, and value formats or ranges}
    \label{tab:detailed_data_items_ch2025}
    \centering 
    \begin{minipage}{\textwidth}
    \begin{tabularx}{\textwidth}{ *{3}{>{\centering\arraybackslash}X} l}
    \toprule
    \textbf{Item} & \textbf{Column} & \textbf{Data type} & \multicolumn{1}{c}{\textbf{Note}}\\ 

    \midrule
    mACStatus & m\_charging& integer& 0: No, 1: Charging\\ \hline
    mActivity & m\_activity & integer & 0: in\_vehicle, 1: on\_bicycle, 2: on\_foot, 3: still, 4: unknown, 5: tilting, 7: walking, 8: running\\ \hline
    mAmbience & m\_ambience& object& List of ambient sound labels and their respective probabilities\\ \hline
    mBle & m\_ble& object& List of bluetooth device address, device\_class, and rssi\\ \hline
    mGps & m\_gps& object& List of (altitude, latitude, longitude, speed)\\ \hline
    mLight & m\_light& float& Ambient light in lx unit\\ \hline
    mScreenStatus & m\_screen\_use& integer & 0: No, 1: Using screen\\ \hline
    mUsageStats & m\_usage\_stats & object & List of app names and their respective usage times (in milliseconds unit)\\ \hline
    mWifi & m\_wifi& object& List of base station ID(bssid) and rssi\\ \hline
    
    wHr & heart\_rate & object & List of heart rate recordings\\ \hline
    wLight & w\_light& float& Ambient light in lx unit\\ \hline
    \raisebox{5ex}[0pt][0pt]{wPedo} &
        \shortstack[c]{%
        burned\_calories \\ distance \\ speed \\ step \\ step\_frequency%
        } &
        \shortstack[c]{%
        float \\ float \\ float \\ integer \\ float%
        } &
        \shortstack[l]{%
        Number of calories \\ Distance in meters \\ Speed in km/h unit \\
        Number of steps \\ Step frequency in a minute%
        } \\ 
    \bottomrule
    \end{tabularx}
    \end{minipage}    
\end{table*}

\begin{figure}[tb!]
  \centering
  \includegraphics[width=1.0\columnwidth]{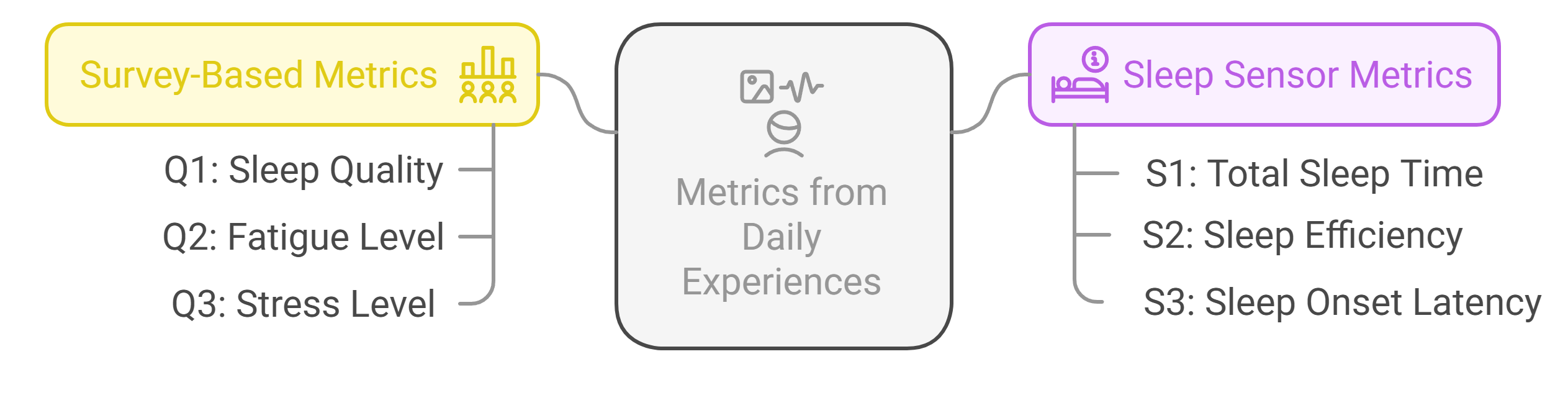}
  \caption{Overview of the six daily metrics related to sleep health, fatigue, and stress, derived from sleep sensor data and self-reported surveys}
  \label{fig:fig_sleep_metrics}
\end{figure}

In addition to the twelve data items measured from smartphones and smartwatches during daytime activities, we also derived the following six daily metrics, which are shown in Figure~\ref{fig:fig_sleep_metrics}, related to sleep health, fatigue, and stress from sleep sensor data and self-reported survey records. The method used to derive these metrics is similar to that presented in our previous work~\cite{oh2024sensor}.

\begin{itemize}
\item \textbf{Q1:} Overall sleep quality as perceived by a subject immediately after waking up.
\item \textbf{Q2:} Physical fatigue of a subject just before sleep.
\item \textbf{Q3:} Stress level of a subject just before sleep.
\item \textbf{S1:} Adherence to the recommended total sleep time.
\item \textbf{S2:} Adherence to the recommended sleep efficiency.
\item \textbf{S3:} Adherence to the recommended sleep onset latency.
\end{itemize}

Specifically, the three survey-based metrics (i.e., Q1, Q2, and Q3) were derived from the individual's pre- and post-sleep questionnaire responses, which were averaged across the entire experimental period. Each questionnaire was originally recorded using a 5-point Likert scale and was converted into a binary format. For instance, the first questionnaire metric (Q1) is assigned a value of \textbf{1} on days when an individual's self-reported sleep quality exceeds their average over the experimental period, and \textbf{0} when it falls below that average. Similarly, the second and third metrics (Q2 and Q3) are assigned a value of \textbf{0} on days when the participant’s fatigue and stress levels, respectively, exceed their average, and a value of \textbf{1} when these levels are below average. In summary, the survey-based metrics are labeled as \textbf{1} when they indicate a subjectively positive outcome based on the individual participant’s self-assessment.

Meanwhile, the three sleep metrics (i.e., S1, S2, and S3), which were derived from the sleep sensor, are designed to reflect the sleep health recommendations established by the National Sleep Foundation~\cite{nsf_guidelines}. These metrics assign a value of \textbf{0} for sleep records that do not meet the recommended guidelines. In contrast, sleep records that satisfy the recommendations are assigned a value of \textbf{1} for S2 and S3, which are binary metrics. Notably, unlike the other metrics, the sleep metric for total sleep time (S1) can take on three distinct values(e.g., \textbf{0}: Not recommended; \textbf{1}: May be appropriate; \textbf{2}: Recommended).

\section{Baseline machine learning model}

\begin{figure}[h!]
  \centering
  \includegraphics[width=0.75\columnwidth]{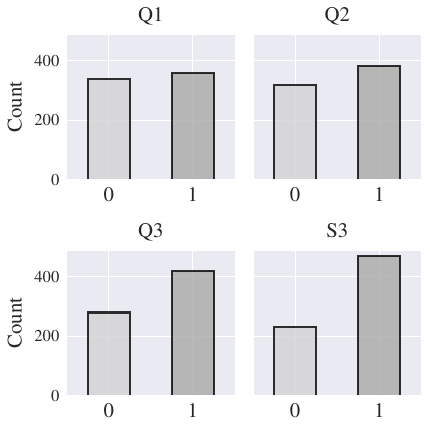}
  \caption{Label distribution (0 and 1) for the four target metrics}
  \label{fig:fig_four_metrics}
\end{figure}

In this section, we present an example of a baseline machine learning model to demonstrate the types of features that can be extracted and the classification performance achievable when these features are input into a model to classify four target metrics: Q1, Q2, Q3, and S3. Figure~\ref{fig:fig_four_metrics} illustrates the number of instances labeled as \textbf{0} and \textbf{1} for these four metrics, highlighting potential class imbalance.

\begin{figure}[b!]
  \centering
  \includegraphics[width=1.0\columnwidth]{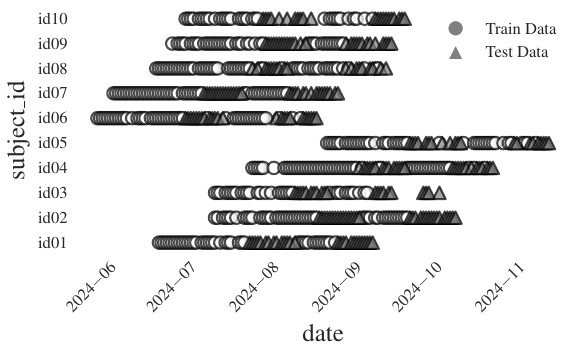}
  \caption{Temporal distribution of training (triangle) and test (circle) data across individual subjects}
  \label{fig:fig_data_split}
\end{figure}

Out of the total 700 days of Lifelog data, 450 days were allocated for training a learning model, while the remaining 250 days were used for evaluation. Since the data were collected from human participants and behavioral changes may occur over extended experimental periods, the test data were not limited to the final phase of participation. Instead, the test data were configured to include both the middle and later stages of each subject’s participation period, thereby better reflecting temporal variations in behavior. It is worth noting that a chi-squared test revealed no significant differences in the distribution of each target metric between the training and test data splits (p$>$ 0.2). Figure~\ref{fig:fig_data_split} shows the temporal distribution of training and test data points for each subject, highlighting the non-contiguous and personalized split across different dates. The discontinuity in daily records is due to the prior exclusion of days that either failed to meet the minimum metric generation criteria (e.g., less than three hours spent in bed) or exhibited excessive missing data.

We derived features from five data sources: wPedo, wHr, wLight, mLight, and mUsage. Given that some data items may be missing in the original dataset, we decided to use descriptive features at fixed time intervals to cope with information loss and reduce the computational burden on the learning model. For wPedo, wHr, wLight, and mLight, we segmented each day into seven time zones (i.e., 00h-06h, 06h-09h, 09h-12h, 12h-15h, 15h-18h, 18h-21h, and 21h-24h) and computed representative statistical values such as the mean, sum, and standard deviation.
\begin{itemize}
\item wPedo: Mean and total step count per time zone
\item wHr: Mean heart rate per time zone and proportion of high heart rate values
\item wLight, mLight: Mean and standard deviation of light intensity per time zone after applying log10 transformation
\end{itemize}
For mUsage, we categorized the most frequently used mobile apps into three broad categories (i.e., System, Social, and Hobby) and computed the total daily usage time in minutes for each category. In addition, four demographic attributes for each subject (i.e., gender, age group [whether aged 40 or older], employment status, and BMI), along with day-of-week information, were included as input features (marked as `DW') for model training.

To assess the baseline model’s performance with the extracted features, we trained classification models using Python 3.12, LightGBM (version 4.6), and scikit-learn (version 1.6). All models were trained using default parameter settings provided by the LightGBM library. Table~\ref{tab:performance_comparison} presents the classification performance of the baseline model for each target metric, measured by macro F1 score. Bolded values indicate the feature combinations that achieved the highest performance in our experiments.

\begin{table}[b!]
    \begin{threeparttable}
    \caption{Performance comparison across different feature combinations}
    \label{tab:performance_comparison}
    \centering
    \begin{minipage}{\textwidth}    
    \begin{tabular}{lccccc}
    \toprule 
    \textbf{Features} (n)& \textbf{Q1} & \textbf{Q2} & \textbf{Q3} & \textbf{S3}\\
    
    \midrule
    wPedo (14) & 0.549 & 0.524 & 0.563 & 0.537 \\
    wPedo+DW (19) & \textbf{0.619} & 0.537 & 0.647 & \textbf{0.651} \\ \hline
    wHr (14) & 0.548 & 0.492 & 0.587 & 0.526 \\
    wHr+DW (19) & 0.527 & 0.568 & 0.577 & 0.539 \\ \hline
    wLight (14) & 0.513 & 0.563 & 0.540 & 0.524 \\
    wLight+DW (19) & 0.557 & 0.565 & 0.604 & 0.594 \\ \hline
    mLight (14) & 0.477 & 0.557 & 0.562 & 0.507 \\
    mLight+DW (19) & 0.579 & 0.574 & 0.592 & 0.641 \\ \hline
    mUsageStats (3) & 0.543 & 0.568 & 0.497 & 0.554 \\
    mUsageStats+DW (8) & 0.562 & \textbf{0.634} & 0.623 & 0.610 \\ \hline
    wPedo+mUsageStats (17) & 0.563 & 0.532 & 0.643 & 0.544 \\
    wPedo+mUsageStats+DW (22) & 0.563 & 0.596 & \textbf{0.674} & 0.575 \\ \hline
    All (59) & 0.531 & 0.552 & 0.635 & 0.603 \\
    All+DW (64) & 0.570 & 0.569 & 0.597 & 0.594 \\
    \bottomrule
    \end{tabular}
    \end{minipage}
    \begin{tablenotes}
    \footnotesize
    \item \textit{Note.} DW means five features about demographics and weekday information.
    \end{tablenotes}
    \end{threeparttable}
\end{table}

\section{Discussion}

The ETRI Lifelog dataset 2024 included more participants (increased from 4 to 10) and a larger number of daily instances (from 220 to 700 days) compared to the 2023 dataset. However, among the seven target metrics defined in 2023~\cite{oh2024sensor}, some were revised in 2024, resulting in a total of six daily metrics. For instance, the Q2 metric, which represented the emotional state just before sleep in 2023, was replaced in 2024 to indicate the fatigue level before sleep. Similarly, the S1 metric for total sleep time, previously a binary classification, was modified into a three-class classification in 2024.

\begin{figure}[t!]
  \centering
  \includegraphics[trim={0 0 35 35}, clip, width=0.65\columnwidth,]{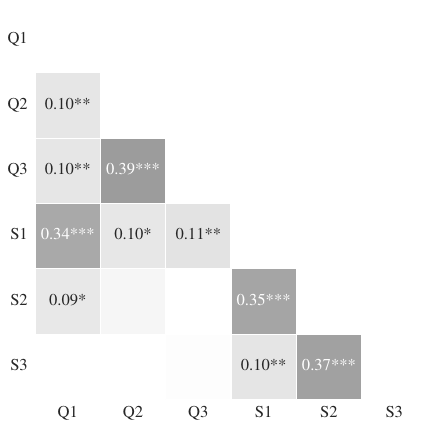}
  \caption{Spearman correlation matrix among six daily metrics. Asterisks indicate significance levels: p$<$0.05 (*), p$<$0.01 (**), and p$<$0.001 (***).}
  \label{fig:fig_correlation_matrix}
\end{figure}

Figure~\ref{fig:fig_correlation_matrix} presents the Spearman correlations among six metrics for the ETRI Lifelog Dataset 2024. Only statistically significant coefficients (p$<$0.05) are shown in the figure. The results indicate that the metrics exhibit weak positive correlations, either directly or indirectly. For example, Q1 is not directly correlated with S3, but it shows a significant correlation with S1, which in turn correlates with S2, and S2 with S3, which implies an indirect association between Q1 and S3. This suggests that the perceived quality of sleep is associated with various behavioral outcomes during sleep. A positive association is also observed between Q2 and Q3. This observation indicates that physical fatigue and stress levels before sleep are positively related.

An analysis of the modeling results presented in Table~\ref{tab:performance_comparison} reveals that the baseline models generally performed better when input features were selectively chosen for each metric, rather than when all available features were used. This finding suggests that machine learning models benefit from a more refined and task-specific feature selection process. These results are consistent with those reported in a previous study conducted last year. For instance, in the cases of Q1 and S3, the highest performance was achieved using features related to step counts in 3-hour time units, while for Q2 and Q3, features based on daily mobile app usage yielded superior results. Additionally, across all baseline models, the inclusion of demographic and weekday features consistently enhanced classification performance.

\begin{figure}[t!]
  \centering
  \includegraphics[width=0.95\columnwidth]{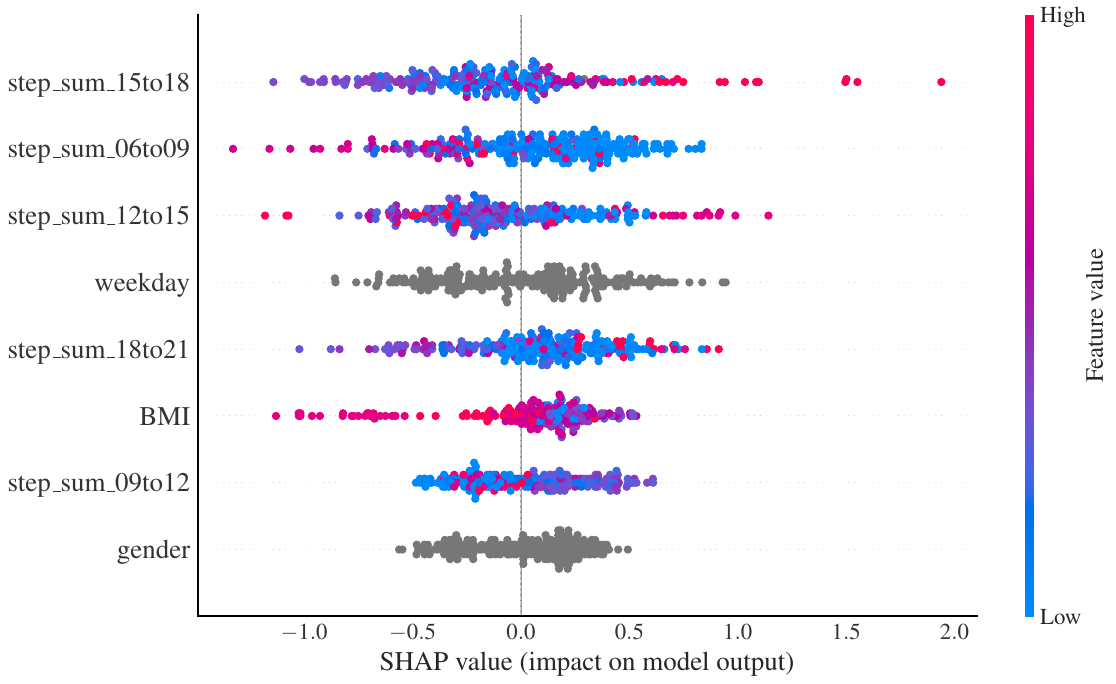}
  \caption{SHAP values for the baseline model predicting Q1 using wPedo+DW features}
  \label{fig:fig_wpedo_q1}
\end{figure}

Figure~\ref{fig:fig_wpedo_q1} presents SHAP value analysis~\cite{lundberg2017unified} for the classification model using `wPedo+DW' features, which showed the best performance in predicting the Q1 metric. The plot displays the top 8 features ranked by their impact on the model's output. The most influential features, in order, are step count from 15:00 to 18:00, step count from 06:00 to 09:00, step count from 12:00 to 15:00, and whether the day is a weekday. Notably, higher step counts between 15:00 and 18:00 contribute positively to predicting Q1 = \textbf{1}, while higher step counts between 06:00 and 09:00 contribute negatively.

\begin{figure}[b!]
  \centering
  \includegraphics[trim={0 16 0 0}, clip, width=0.90\columnwidth]{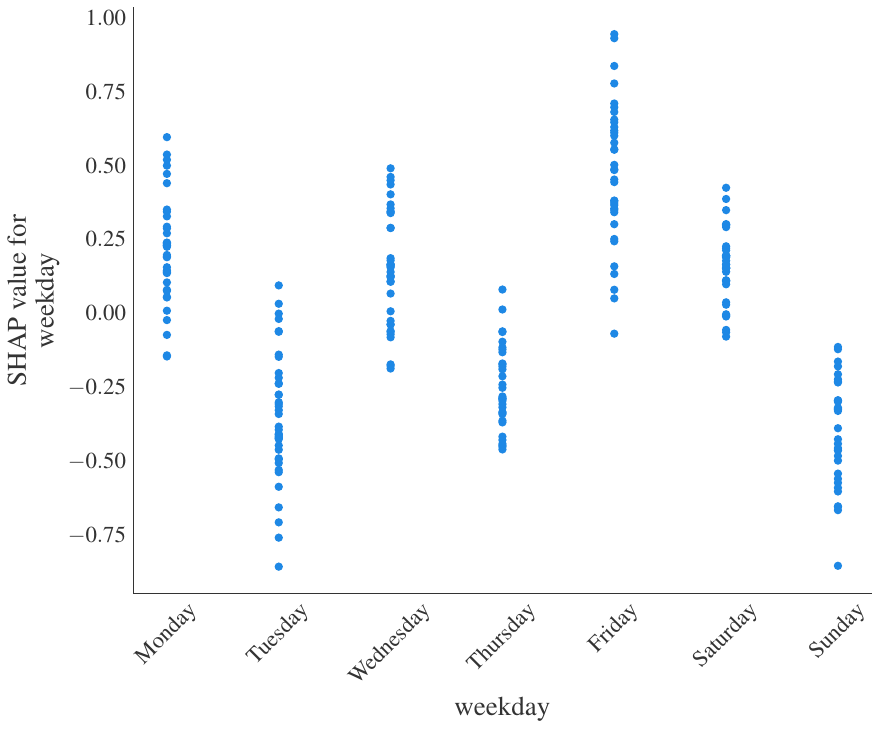}
  \caption{SHAP values by day of the week (weekday feature)}
  \label{fig:fig_wpedo_q1_shap_weekday}
\end{figure}

Figure~\ref{fig:fig_wpedo_q1_shap_weekday} illustrates the SHAP values for the weekday feature across different days of the week. The SHAP values tend to be higher on Fridays and lower on Sundays and Tuesdays, which aligns with general intuition. This suggests that, in addition to sensor-derived features, contextual factors related to participants’ daily environments play a role in model predictions. It also helps explain the performance improvement observed in all models when `DW' features (i.e., demographics and weekday information) were included as input, as previously shown in Table~\ref{tab:performance_comparison}.

\section{Conclusion}
In this paper, we detailed the main components and characteristics of the ETRI Lifelog Dataset 2024. We also presented a baseline machine learning model for sleep quality prediction along with an example analysis of feature importance using SHAP values. This dataset has been made publicly available for research and has served as a key resource for the fourth Human Understanding AI Paper Challenge, held in conjunction with ICTC 2025. We anticipate that this new dataset will likewise provide a valuable foundation for a wide range of studies, including those on physical activity, sleep quality, stress patterns, and broader aspects of daily human life.

\section*{Acknowledgment}
This work was supported by Electronics and Telecommunications Research Institute (ETRI) grant funded by the Korean government. [25ZB1100, Core Technology Research for Self-Improving Integrated Artificial Intelligence System].

\vspace{12pt}

\end{document}